# ARCHISOUND: AUDIO GENERATION WITH DIFFUSION

FLAVIO SCHNEIDER

Master's Thesis

Supervised by Zhijing Jin, Prof. Bernhard Schölkopf

ETH Zurich
January 2023

# ABSTRACT

The recent surge in popularity of diffusion models for image generation has brought new attention to the potential of these models in other areas of media generation. One area that has yet to be fully explored is the application of diffusion models to audio generation. Audio generation requires an understanding of multiple aspects, such as the temporal dimension, long term structure, multiple layers of overlapping sounds, and the nuances that only trained listeners can detect. In this work, we investigate the potential of diffusion models for audio generation. We propose a set of models to tackle multiple aspects, including a new method for text-conditional latent audio diffusion with stacked 1D U-Nets, that can generate multiple minutes of music from a textual description. For each model, we make an effort to maintain reasonable inference speed, targeting real-time on a single consumer GPU. In addition to trained models, we provide a collection of open source libraries with the hope of simplifying future work in the field. Samples can be found at `bit.ly/audio-diffusion`.

# contents

# INTRODUCTION

Music is an art of time at the intersection of fine-grained perception and symbolic patter recognition. In this work, we will investigate the use of diffusion model to generate music, or more broadly audio, in order to gain a deeper understanding of this intersection using modern deep learning diffusion models.

## 1.1 AUDIO GENERATION

Audio generation refers to the process of automatically synthesizing novel waveforms using deep learning models. Audio generation has been commonly approached in two different ways: symbolically or at the waveform level. Symbolically generating audio involves creating a representation of the audio using symbols, such as MIDI data, which can then be converted into an audio waveform. This method is often easier to work with, but it can be difficult to capture all the nuanced details of a sound using symbols. Waveform-based audio generation, on the other hand, involves generating the raw audio waveform directly. This method is more complex, due to the sheer amount of values that have to be generated per second, but it allows for a more precise and detailed representation of sound, that includes all of its intricacies. Furthermore, audio generation can be unconditional or conditional. Unconditional models are trained only on audio data and are able to generate new samples without any additional input. Conditional models, on the other hand, are trained on pairs of audio data and some kind of conditioning information, such as a text description, genre label, lyrics, speaker id, or some other description of the audio. At inference time, this conditioning information can be used to guide the generation of novel audio samples that match the desired characteristics. In this thesis, we will explore methods of conditional and unconditional waveform-level generation.

## 1.2 CHALLENGES

Multiple tradeoffs have to be considered when generating audio at the waveform level. To generate a single second of high quality 48kHz stereo audio, 96000 values must be generated, which is comparable in size to a medium resolution image. If the goal is to generate an entire song (hundreds of seconds) maintaining high-*quality* and a reasonable generation *speed*, this task becomes much more challenging. A common approach to generating long audio sequences is to do so in chunks, however, if the *context length*, or the amount of audio that the model can consider at any given time is not sufficient, the resulting structure may not be consistent over multiple seconds or minutes of generation. A longer context may allow for more consistent coarse structure, but may also lead to lower overall quality of detail or vice-versa.

## 1.3 EXISTING METHODS

In this section, we will review some of the most well-known or influential waveform-based methods that have been developed to date.

One of the pioneering waveform level generation models is WaveNet (2016 [8]), a fully convolutional architecture that exploits dilated convolutions with various dilation factors in order to capture a large context. It's able to synthesize a few seconds of both speech and classical piano music at 16kHz. Jukebox (2020 [2]) uses multiple quantized autoencoders to discretize sounds at 3 different resolutions, followed by a cascade of transformer upsampler models to generate the quantized representations autoregressively. Jukebox is able to generate 44kHz music conditioned on lyrics, artists and genres. The stack of transformers trades-off generation speed for structure and quality. AudioLM (2022 [1]) uses a (residual) quantized autoencoder to compress the waveform into discrete tokens and a semantic encoder, later a cascade of transformer decoders (semantic, coarse, fine) is used to generate 16kHz audio continuations top-down from the semantic representation. Musika (2022) trains a set of 1D convolutional autoencoders to compress log-magnitude spectrograms, and a vocoder to reconstruct both phase and magnitude from the compressed representation, using a 2D GAN discriminator trained on sequential chunks of audio exploits this process autoregressively to generate longer sequences of 44kHz audio. This method has a limited context length, but is very efficient given the 1D structure of convolutions. Riffusion[1] (2022) fine-tunes the Stable Diffusion model [12] on chunks of mel-spectrograms of 5s at 44kHz, and uses style transfer to generate multiple coherent concatenated images while conditioning on a textual description of the song. This method has a limited 5s context length, and trades off speed given the large 2D architecture, but works surprisingly well considering that the original model is trained on images, not audio.

## 1.4 RESEARCH QUESTIONS

Diffusion models have recently demonstrated exceptional capabilities in the field of image generation [11, 12], leading to an explosion of incredible AI generated art [2]. Iteratively removing small amounts of noise from pure noise allows diffusion models to hallucinate novel samples that have common attributes to the data in the training set. Compared to GANs, diffusion models in the image domain don't suffer from training instability, scale well with parameter size, and have good mode coverage.

---

[1] https://www.riffusion.com/about

[2] https://www.midjourney.com/showcase/

As long as the training data can be progressively corrupted from a clean to a fully covered state, diffusion models have the potential to be applied to multiple domains to generate novel samples. This opens up a wide range of possibilities beyond image generation, including video and audio generation.

In this thesis, we explore the potential of diffusion models for audio generation. We will explore whether diffusion models can be used on audio as effectively as with images. The aim is to generate high-quality 48kHz stereo audio as efficiently as possible and to control the generation in different ways, with a focus on text-conditional audio generation.

1.5 CONTRIBUTIONS

1.5.1 *Models*

We introduce the following models, some of which are/will be accessible in the `archisound` library:

- *Long*: a latent diffusion model for *text-conditional music generation* that is capable of generating audio with an extended context of multiple minutes at 48kHz, targeting context length and structure (~857M parameters).

- *Crisp*: a text-conditional audio generation diffusion model with a context of tens of seconds at 48kHz, targeting simplicity and high-quality waveforms (~419M parameters).

- *Upsampler*: a diffusion model to uspsample music from 3kHz to 48kHz (~238M parameters).

- *Vocoder*: A diffusion model to reconstruct 48kHz waveforms from 80-channel mel-spectrograms, variable input length (~178M parameters).

1.5.2 *Libraries*

Moreover, we open-source the following libraries, on which previous models are based:

- `archisound`[3], our library including trained models ready to use. This repository doesn't contain any modelling code, but acts as a wrapper and documentation for our models hosted on Huggingface [4].

- `audio-diffusion-pytorch`[5] (ADP), the main library including the proposed audio diffusion models. This library has both `a-unet` and `audio-encoders-pytorch` as dependencies. At the time of writing, this library has 550+ stars on GitHub, and has been downloaded more than 50000 times on pip.

- `a-unet`[6], a highly customizable library to build U-Net architectures in any dimension, expansible with multiple blocks and plugins. This library can be used for any type of grid data: 1D, 2D, 3D.

- `audio-encoders-pytorch`[7] (AEP), a set of encoders and autoencoders for 1D data.

---

3 https://github.com/archinetai/archisound
4 https://huggingface.co/archinetai
5 https://github.com/archinetai/audio-diffusion-pytorch
6 https://github.com/archinetai/a-unet
7 https://github.com/archinetai/audio-encoders-pytorch

Some additional libraries we open-soruce that are not documented in this thesis, but might nevertheless be interesting to the reader, include: `cqt-pytorch`[8] for invertible CQT spectrograms using NSGT, and `bitcodes-pytorch`[9] a method for vector-quantization into binary codes.

## 1.6 STRUCTURE OF THE THESIS

In Chapter 2, we present the various audio representations and provide a set of tradeoffs that must be considered when selecting an appropriate representation. In Chapter 3, we describe the general principles of diffusion and then delve into the specific diffusion methods that we have tested. In Chapter 4, we examine our custom architectures, including the U-Net and autoencoder, and provide detailed descriptions of each component and how they can be easily integrated into our library. In Chapter 5, we propose a range of diffusion models that combine the diffusion methods from Chapter 3 with our custom architecture from Chapter 4. Finally, in Chapters 6 and 7, we discuss potential future work and present our conclusions.

---

8 https://github.com/archinetai/cqt-pytorch
9 https://github.com/archinetai/bitcodes-pytorch

# AUDIO REPRESENTATION

In the following section, we will introduce the different types of audio representation that we can choose from, and compare the different tradeoffs. Before that, we'll have a look at the different desirable properties that should be considered.

## 2.1 DESIRABLE PROPERTIES

### 2.1.1 *Compressibility*

We define compressibility as the approximate number of values per second needed for high-quality audio compared to the original waveform, and how many can be easily removed without a significant loss in fidelity, e.g. by applying a convolutional only autoencoder on the representation.

#### 2.1.1.1 *Perceptibility*

Perceptibility implies how close is the representation to human hearing, this part is important since if we are compressing a representation that carries a lot of information we are not able to perceive in the first place we will lose a lot of useful capacity. More specifically, humans hear sound in the range of frequency from 20Hz to 20kHz, on a logarithmic scale, which means that the frequency resolution decreases as we approach 20kHz.

### 2.1.2 *Decodability*

Decodability refers to how simple and fast is to decode the given representation back to the waveform domain that can be reproduced.

### 2.1.3 *Diffuseability*

Diffusability is a set of desirable properties that are important in order for a diffusion model to be applicable. In particular, (1) the values should be approximately in the range $[-1, 1]$, (2) the signal should ideally have some inductive biases that can be exploited by the network (primarily 1D or 2D convolutional blocks), (3) time-shift invarance if we are doing inpainting or autoregressive generation, i.e. the representation should look the same at different time steps for the same sound, (4) the values should not have too many values in the time dimension, as that would slow down the diffusion process significantly.

## 2.2 WAVEFORM

Waveforms are the default representation and describe audio as movement in air pressure, the tensor **x** used to represent a waveform has shape $[C, T]$ where C is the number of channels (C = 2 for stereo audio), and T the number of points used to represent the audio, e.g. if T = 48000 and we are representing audio at 48 kHz, then the tensor will contain 1 second of audio. If we want to generate multiple seconds/minutes of high-quality audio, the tensor sizes will be very large. As a good comparison, the number of values of a standard image of shape 256x256x3 will correspond approximately 2 seconds of stereo audio (48000x2). We observed that with a standard 1D convolutional autoencoder, we can get 2x compression of a waveform without significant loss in quality. The advantage of using waveforms is that no decoding is involved and it's easily diffuseable for short sequences, the disadvantages are that long sequences are slow to diffuse directly, and that waveforms are a basic representation which doesn't consider perceptibility. In fact, if we apply a standard L1 or L2 loss function on waveforms, high frequencies will be harder to reconstruct than low frequencies. High frequencies tend to vary more rapidly over time, hence they will typically have larger differences between the two waveforms.

## 2.3 SPECTROGRAMS

### 2.3.1 *STFT*

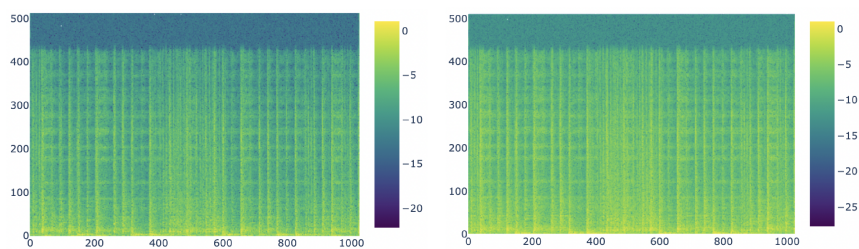

Figure 1: Real and imaginay parts of a single channel STFT, scaled with $\log(|\cdot|)$ for visibility.

Short-Time Fourier-Transform (STFT) are a common way to represent audio into a complex-valued tensor of shape $[C, F, L]$ where C is the number of channels, F is the number of frequencies and L is the length. Each column represents a small chunk of the original wave-form, where the y-axis represents linearly scaled frequencies, and the x-axis represents time. The STFT at a point $(f, l)$ is defined as:

$$\text{STFT}\{\mathbf{x}\}(f, l) = \mathbf{X}(f, l) \tag{1}$$

$$:= \sum_{t=0}^{T} \mathbf{x}_t \cdot \mathbf{w}_{t-l} \cdot \exp\left(-i\frac{2\pi t}{T}f\right) \tag{2}$$

$$= \sum_{t=0}^{T} \mathbf{x}_t \cdot \mathbf{w}_{t-l} \cdot \left(\cos\left(\frac{2\pi t}{T}f\right) - i\sin\left(\frac{2\pi t}{T}f\right)\right) \tag{3}$$

Where $\mathbf{w}$ is a window function of shape $W$ that is used to isolate the chunk of interest from the signal. Usually, the Hann window is used as $\mathbf{w}$. Interestingly, we can consider the STFT as 1D-convolution with predefined filters composed of cosine waves for the real part, and sine waves for the imaginary one, making the real/imaginary STFT output stacks of $L$ channels each, as you would obtain if convolving the signal with $L$ kernels of length $W$. The STFT can be computed efficiently using the Fast Fourier Transform (FFT), and can also be easily inverted.

Instead of real and imaginary parts, often the STFT is converted into magnitude and phase, a representation that can be inverted back to real and imaginary parts, and hence back to wavefrom. Magnitude and phase are defined as $\text{mag}(\mathbf{X}(f, l)) := |\mathbf{X}(f, l)|^2$ and $\text{phs}(\mathbf{X}(f, l)) := \arctan\left(\frac{\Im(\mathbf{X}(f,l))}{\Re(\mathbf{X}(f,l))}\right)$ as shown in Figure 2.

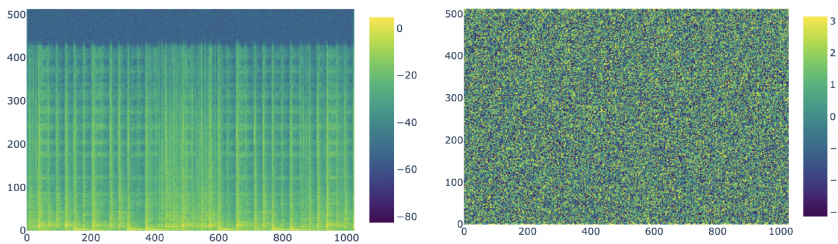

Figure 2: Magnitude spectrogram and phase of a single channel STFT, magnitude scaled with $\log(\cdot)$ for visibility.

We observed that with a standard 1D convolutional autoencoder magnitude spectrogram can be easily compressed up to 32x, with very small loss in quality, on the other hand phase is very hard to compress. This intuitively make sense given that phase looks random and hence the inductive biases of spatial locality that convolution exploit do not easily apply. A common practice is to discard the phase altogether, and to train an additional model (called Vocoder) to predict the phase or directly the wavefrom from the magnitude information. Iterative algorithm such as Griffin-Lim can also be used to recover phase, but are known to produce artifacts. Magnitude-only spectrograms favor compressibility, but trade off an easily decodeable representation.

Compressing real and imaginary parts directly is also challenging, since a lot of randomness is still present in the details of the grid. In a sense, converting the transform to magnitude and phase disentangles the hard to compress from the easy to compress parts.

In terms of preceptibility, magnitude spectrograms are more interesting since each frequency has the same importance, however humans distinguish frequencies on a logarithmic scale, which is still not properly represented with linear-frequency magnitude spectrograms.

If properly normalized, we can diffuse directly either the real and imaginary part, or magnitude spectrograms. However, similarly to compression, we found that phase is harder to diffuse directly.

## 2.3.2 MEL

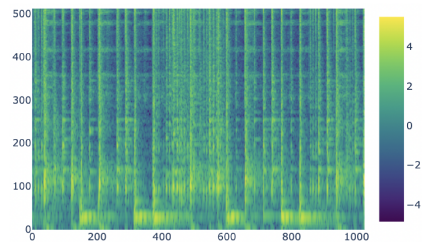

Figure 3: MEL-scale spectrogram, magnitude scaled with $\log(\cdot)$ for visibility.

Mel spectrograms are obtained by scaling magnitude spectrograms by $2595 \log_{10}(1 + \frac{f}{700})$, the mel scale, this special logarithmic scale is based on the perceived pitch of a sound, rather than its actual frequency. Hence, the information it provides is closer to human perception than a linearly scaled spectrogram, making it a very common choice in audio applications. The disadvantage of the mel scale is that it's not invertible back to the linear scale without loss of information. To recover the linearly scaled magnitude spectrograms, optimization based methods are used, or trained vocoders that act both as scale inversion and phase reconstruction in one go.

# EXISTING DIFFUSION METHODS

Diffusion models, first proposed in [3, 17] are most commonly implemented with U-Net [7, 13] that is repeatedly called during inference for each denoising step. Since the same network is called multiple times during sampling, the weights are shared, making it a recurrent model. Since the data can be progressively corrupted from a clean to a fully covered state, we can use this trick to jump to any intermediate noise level and denoise a single step, backpropagating only once during training. From the perspective of recurrent models, (forward) diffusion allows us to recover the memory at an intermediate state (which we can see as the corrupted datapoint) without the need to backpropagate the entire chain. This is a useful technique for efficiently generating intermediate states, and has the advantage that it can be highly parallelized during training. Compared to recurrent models, the memory state is predefined by the (noise) corruption process and not fully learned. Diffusion exploits very similar principles as autoregressive transformer models [19], namely a highly parallelizeable training process and repeated network call with weight-sharing during sampling. Compared to other generative models like GANs, diffusion models are easier to train and don't suffer from instability problems arsing from having to coordinate a generator and discriminator.

Diffusion models are a category of powerful generative models first introduced in [17] (2015), and later popularized in [3] (2020), thanks to the impressive results obtained in image generation on CIFAR10.

In this section, we will examine different diffusion methods. First, the seminal DDPM [3] method, which involves training the diffusion process with a finite number of denoising steps. Following that, DDIM [18] introduces a few changes that generalize DDPM to an arbitrary number of steps. Then we will introduce V-diffusion from [16], a continuous diffusion method that aims to improve the mixing of the signal-to-noise ratio from DDIM. For DDPM and V-diffusion, we will highlight the most important operations, namely: (1) noising the original datapoint (signal) to a desired noise, (2) denoising a single step with the use of our (trained) network, (3) the training objective used, and (4) a sampling technique that repeatedly applies (2).

## 3.1 DDPM-DIFFUSION

DDPM [3] is one of the seminal works in diffusion models. The method starts by assuming that $\mathbf{x}_0^{(0)}, \ldots, \mathbf{x}_0^{(D)}$ is a dataset of D i.i.d. points sampled from an unknown distribution $q(\mathbf{x}_0)$, (the subscript indicates the noise level from 0 to a maximum of T) and that $q(\mathbf{x}_t \mid \mathbf{x}_{t-1}) := \mathcal{N}(\mathbf{x}_t \mid \boldsymbol{\mu}_t = \sqrt{1-\beta_t}\mathbf{x}_{t-1}, \Sigma_t = \beta_t I)$. In words, if we want to increase the noise level of our datapoint $\mathbf{x}_{t-1}$ by one step to level t, we'll have to sample a normal distribution with the mean and convariance dependent on the previous point and some hyperparameters $\beta_1, \ldots, \beta_T$ called *variance schedule*, which control the increase in noise level from the previous point.

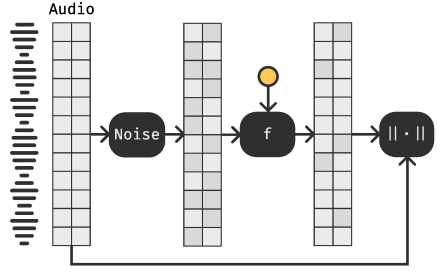

Figure 4: Diffusion training.

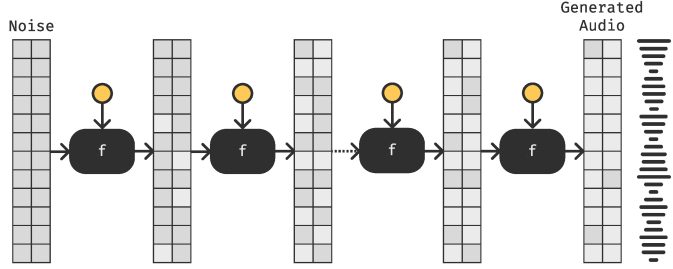

Figure 5: Diffusion inference.

### 3.1.1 *Noising (0 → t)*

By using the previous assumptions, we can derive $q(\mathbf{x}_t \mid \mathbf{x}_0)$, i.e. a way to directly jump from noise level 0, (our clean datapoint) to noise level t. This procedure is called the forward diffusion process. Using the reparemetrization trick, it can be shown that this is also a normal distribution, formulated as:

$$q(\mathbf{x}_t \mid \mathbf{x}_0) := \mathcal{N}\left(\mathbf{x}_t \mid \boldsymbol{\mu}_t := \sqrt{\bar{\beta}_t}\mathbf{x}_0, \Sigma_t := (1-\bar{\beta}_t)I\right) \quad (4)$$

where $\bar{\beta}_t := \prod_{s=1}^{t}(1-\beta_t)$ depends on all $\beta_t$ selected in $q(\mathbf{x}_t \mid \mathbf{x}_{t-1})$. Since this is a normal distribution, with the mean vector and standard deviation we can easily sample $\mathbf{x}_t$, the noisy version of $\mathbf{x}_0$, as:

$$\mathbf{x}_t = \sqrt{\bar{\beta}_t}\mathbf{x}_0 + \sqrt{1-\bar{\beta}_t}\boldsymbol{\epsilon}_t \quad (5)$$

Where $\boldsymbol{\epsilon}_t \sim \mathcal{N}(0, I)$.

### 3.1.2 Denoising (t − 1 ← t)

The reverse process distribution $q(\mathbf{x}_{t-1} \mid \mathbf{x}_t)$ is also a normal distribution. However, it cannot be directly estimated as it depends on the entire dataset. Instead, we train a neural network with parameters $\theta$ as an approximation:

$$p_\theta(\mathbf{x}_{t-1} \mid \mathbf{x}_t) := \mathcal{N}(\mathbf{x}_{t-1} \mid \boldsymbol{\mu}_\theta(\mathbf{x}_t), \Sigma_\theta(\mathbf{x}_t)) \tag{6}$$

If our model is trained properly, similarly to the forward process, we will be able to carry out a single denoising step by sampling the normal distribution using the learned mean and variance.

### 3.1.3 Training Objective

To train our model, we need a handle on the true mean and covariance of the reverse process $q(\mathbf{x}_{t-1} \mid \mathbf{x}_t)$. As we have seen before, this is not directly tractable, however, if we include additional information about either $\mathbf{x}_0$ (the true data point), or $\boldsymbol{\epsilon}_t$ (the noise used to get $\mathbf{x}_t$ from $\mathbf{x}_0$ in the forward process) we can compute a different but tractable auxiliary distribution. In the case where $\mathbf{x}_0$ is given, the distribution is:

$$q(\mathbf{x}_{t-1} \mid \mathbf{x}_t, \mathbf{x}_0) = \mathcal{N}\left(\mathbf{x}_{t-1} \mid \tilde{\boldsymbol{\mu}}(\mathbf{x}_t, \mathbf{x}_0), \tilde{\Sigma}(\mathbf{x}_t, \mathbf{x}_0)\right) \tag{7}$$

With mean $\tilde{\boldsymbol{\mu}}(\mathbf{x}_t, \mathbf{x}_0) := \frac{\sqrt{1-\beta_t}(1-\bar{\beta}_{t-1})}{1-\bar{\beta}_t}\mathbf{x}_t + \frac{\sqrt{\bar{\beta}_{t-1}}\beta_t}{1-\bar{\beta}_t}\mathbf{x}_0$ and covariance $\tilde{\Sigma}(\mathbf{x}_t, \mathbf{x}_0) := \frac{1-\bar{\beta}_{t-1}}{1-\bar{\beta}_t}\beta_t I$, as shown in [3]. To train our network, we will then minimize the divergence between this tractable distribution and the distribution estimated with our model:

$$L_t := D_{KL}\left[q(\mathbf{x}_{t-1} \mid \mathbf{x}_t, \mathbf{x}_0) \parallel p_\theta(\mathbf{x}_{t-1} \mid \mathbf{x}_t)\right] \tag{8}$$

$$= \mathbb{E}_{\mathbf{x}_0}\left[\frac{1}{2\|\Sigma_\theta(\mathbf{x}_t)\|_2^2}\|\tilde{\boldsymbol{\mu}}(\mathbf{x}_t, \mathbf{x}_0) - \boldsymbol{\mu}_\theta(\mathbf{x}_t)\|_2^2\right] \tag{9}$$

Which amounts to a simple L2 loss between the auxiliary mean, and the true mean estimated by the model, with some extra scaling factor that is dependent on the covariance, in [3] the covariance is fixed to $\Sigma_\theta(\mathbf{x}_t) = \beta_t I$. A more rigorous argument using variational inference can be applied to show that this is a lower bound of the negative log-likelihood of the data distribution. More concretely, our model $f_\theta$ will output an estimated mean given the noisy datapoint and the noise level as input: $\boldsymbol{\mu}_\theta(\mathbf{x}_t) = f_\theta(\mathbf{x}_t, t)$, which we can then use to sample the next $\mathbf{x}_{t-1}$ from a normal distribution.

If instead we assume $\boldsymbol{\epsilon}_t$ is given, we can follow a similar procedure to get the loss $L_t$:

$$L_t := D_{KL}\left[q(\mathbf{x}_{t-1} \mid \mathbf{x}_t, \boldsymbol{\epsilon}_t) \parallel p_\theta(\mathbf{x}_{t-1} \mid \mathbf{x}_t)\right] \tag{10}$$

$$= \mathbb{E}\left[\frac{\beta_t^2}{2\beta_t(1-\bar{\beta}_t)\|\Sigma_\theta(\mathbf{x}_t)\|_2^2}\|\boldsymbol{\epsilon}_t - \boldsymbol{\epsilon}_\theta(\mathbf{x}_t)\|_2^2\right] \tag{11}$$

In this case our model will estimate the noise instead of the mean of the datapoint $\mathbf{x}_t$, i.e. $\boldsymbol{\epsilon}_\theta(\mathbf{x}_t) = f_\theta(\mathbf{x}_t, t)$, however we can still recover the mean as: $\tilde{\boldsymbol{\mu}} = \frac{1}{\sqrt{1-\beta_t}}\left(\mathbf{x}_t - \frac{\beta_t}{\sqrt{1-\bar{\beta}_t}}\boldsymbol{\epsilon}_t\right)$. Empirically, it has been shown in [3] that the objective can be simplified further by ignoring the scaling factor:

$$L_t = \mathbb{E}_{\boldsymbol{\epsilon}_t}\left[\|\boldsymbol{\epsilon}_t - \boldsymbol{\epsilon}_\theta(\mathbf{x}_t)\|_2^2\right] \tag{12}$$

The final objective function to train the model is then computed with random noise levels t sampled from a uniform distribution.

$$L := \mathbb{E}_{t \sim [1,T]}[L_t] \tag{13}$$

### 3.1.4 *Sampling*

Sampling in DDPM is very straightforward, we start with $\mathbf{x}_T \sim \mathcal{N}(0, I)$ and recursively call the model T times using at each step the estimated means $\boldsymbol{\mu}_\theta(\mathbf{x}_t)$ (or noises $\boldsymbol{\epsilon}_\theta(\mathbf{x}_t)$) of the T normal distributions to get each subsequent sample: $\mathbf{x}_{T-1} \sim p_\theta(\cdot \mid \mathbf{x}_T), \ldots, \mathbf{x}_1 \sim p_\theta(\cdot \mid \mathbf{x}_2)$, $\mathbf{x}_0 \sim p_\theta(\cdot \mid \mathbf{x}_1)$ where $\mathbf{x}_0$ will be our generated output data point. Note that this is a stochastic sampling process, since at each step additional noise is added from sampling the normal distribution.

### 3.1.5 *Limitations*

This method requires on the order of hundreds of sampling steps to get good quality samples. Compared to more modern methods that follow, the number of steps T is a fixed hyperparameter both during training and sampling, limiting its flexibility.

## 3.2 DDIM

DDIM [18], is another seminal work for diffusion models. By introducing a few changes to DDPM, the number of sampling steps used during inference can be dynamically changed while maintaining the same training procedure. This allows to sample between x10 and x100 faster, and to trade speed for quality at will. A direct implication of having a variable number of steps during sampling is that we can train with very large T, or even infinitely large T, leading to a continuous diffusion process. The idea of DDIM is that if we know both $\mathbf{x}_0$ and $\mathbf{x}_t$, we can use $q(\mathbf{x}_{t-1} \mid \mathbf{x}_t, \mathbf{x}_0)$ to sample $\mathbf{x}_{t-1}$. There are two possibilities, either train our network to predict directly (i.e. no sampling) $\mathbf{x}_0$, or train our network to predict the noise $\boldsymbol{\epsilon}_t$ (as done in DDPM) that combined with $\mathbf{x}_t$ can be used to infer $\mathbf{x}_0$. A key observation is that using this alternative method doesn't change the training objective, as the objective only depends on the backward diffusion process.

Importantly, we can use a different forward process to recover the next step, for example use $q(\mathbf{x}_{t-2} \mid \mathbf{x}_t, \mathbf{x}_0)$ to jump directly to $\mathbf{x}_{t-2}$ instead of $\mathbf{x}_{t-1}$, essentially skipping a sampling step and speeding up the process. If we make the time-step continuous, we can jump to any intermediate step in $(0, t]$. Even more interestingly, this continuous sampling procedure can be viewed from the lens of ordinary differential equations, allowing us to use a variety of existing samplers, like the basic Euler methods or more advanced ODE samplers.

## 3.3 V-DIFFUSION

V-diffusion, or v-objective diffusion [16], is a diffusion method inspired from DDIM, trained with a continuous value $\sigma_t \in [0, 1]$. This is the method we found to work best on a variety of audio tasks. In v-diffusion, if $\sigma_t = 0$ then $\mathbf{x}_{\sigma_t}$ represents a data point $\mathbf{x}$ from the data distribution and if $\sigma_t = 1$, it will be Gaussian noise $\boldsymbol{\epsilon}$. In DDIM we can choose to either use the model to predict $\mathbf{x}_0$, or use it to predict $\boldsymbol{\epsilon}_t$, in this case however, a velocity value $\mathbf{v}_{\sigma_t}$ is estimated from which both $\mathbf{x}_0$ and $\boldsymbol{\epsilon}_{\sigma_t}$ can be inferred.

### 3.3.1 Noising ($0 \to \sigma_t$)

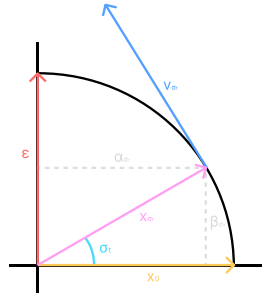

Figure 6: V-Diffusion semicircle

The noising process uses a weighting on a circle:

$$\mathbf{x}_{\sigma_t} = \alpha_{\sigma_t}\mathbf{x}_0 + \beta_{\sigma_t}\boldsymbol{\epsilon} \tag{14}$$

Where $\alpha_{\sigma_t} := \cos(\phi_t)$, and $\beta_{\sigma_t} := \sin(\phi_t)$, where $\phi_t := \frac{\pi}{2}\sigma_t$. When $\sigma_t = 0$, then $\mathbf{x}_{\sigma_t} = \mathbf{x}_0$, i.e. no noise is added, if instead $\sigma_t = 1$, then $\mathbf{x}_{\sigma_t} = \mathbf{x}_1 = \boldsymbol{\epsilon}$, i.e. only noise $\boldsymbol{\epsilon} \sim \mathcal{N}(0, I)$. Intuitively, using the weighting on a circle makes sure that as we move $\sigma_t$ linearly from 0 to 1 the noising process slowly removes information from $\mathbf{x}_0$. By sampling a random $\sigma_t \in [0, 1]$, we are more likely to pick a value that resembles $\mathbf{x}_0$ instead of pure noise $\boldsymbol{\epsilon}$, meaning that the model will more often see data with smaller amount of noise. Empirically, this has been shown to be beneficial over standard DDIM diffusion.

### 3.3.2 Denoising ($\sigma_{t-1} \leftarrow \sigma_t$)

To denoise a from noise level $\sigma_t$ to noise level $\sigma_{t-1}$, we can use our velocity-estimating model $\hat{\mathbf{v}}_{\sigma_t} = f_\theta(\mathbf{x}_{\sigma_t}, \sigma_t)$, note that the velocity here is defined as the derivative $\mathbf{v}_{\sigma_t} := \frac{\partial \mathbf{x}_{\sigma_t}}{\sigma_t}$, i.e. how much does the datapoint change with a small change in the noise level $\sigma_t$ (see circle figure). As mentioned before, using an estimate of $\mathbf{v}_t$, we can obtain both $\mathbf{x}_0$ and $\boldsymbol{\epsilon}_t$, which in turn can be used to estimate $\mathbf{x}_{\sigma_{t-1}}$ in DDIM style:

$$\hat{\mathbf{v}}_{\sigma_t} = f_\theta(\mathbf{x}_{\sigma_t}, \sigma_t) \tag{15}$$

$$\hat{\mathbf{x}}_0 = \alpha_{\sigma_t}\mathbf{x}_{\sigma_t} - \beta_{\sigma_t}\hat{\mathbf{v}}_{\sigma_t} \tag{16}$$

$$\hat{\boldsymbol{\epsilon}}_{\sigma_t} = \beta_{\sigma_t}\mathbf{x}_{\sigma_t} + \alpha_{\sigma_t}\hat{\mathbf{v}}_{\sigma_t} \tag{17}$$

$$\hat{\mathbf{x}}_{\sigma_{t-1}} = \alpha_{\sigma_{t-1}}\hat{\mathbf{x}}_0 + \beta_{\sigma_{t-1}}\hat{\boldsymbol{\epsilon}}_t \tag{18}$$

In the previous equations, the first 3 lines show how to recover the clean datapoint $\mathbf{x}_0$ and the noise $\boldsymbol{\epsilon}_t$ from $\mathbf{v}_t$, and the last line, remixes the noise with the initial datapoint to get $\mathbf{x}_{\sigma_{t-1}}$. The previous equations can be formally obtained by using trigonometric properties on the definition of velocity (as shown in the appendix of [16]), and intuitively understood by rearranging vectors on the semicircle.

### 3.3.3 Training Objective

By taking the derivative of the noising formulation, we can compute the true velocity $\mathbf{v}_{\sigma_t} = \alpha_{\sigma_t}\boldsymbol{\epsilon} - \beta_{\sigma_t}\mathbf{x}_{\sigma_t}$. The training objective is then:

$$L = \mathbb{E}_{t \sim [0,1], \sigma_t}\left[\|\hat{\mathbf{v}}_{\sigma_t} - \mathbf{v}_{\sigma_t}\|_2^2\right] \tag{19}$$

$$= \mathbb{E}_{t \sim [0,1], \sigma_t}\left[\|f_\theta(\mathbf{x}_{\sigma_t}, \sigma_t) - \alpha_{\sigma_t}\boldsymbol{\epsilon} - \beta_{\sigma_t}\mathbf{x}_{\sigma_t}\|_2^2\right] \tag{20}$$

$$\tag{21}$$

Where $\boldsymbol{\epsilon} \sim \mathcal{N}(0, I)$ and $\mathbf{x}_{\sigma_t}$ is computed according to the noising formulation.

### 3.3.4 Sampling ($\sigma_0 = 0 \leftarrow \cdots \leftarrow \sigma_{t-1} \leftarrow \sigma_t = 1$)

To obtain a new data point $\hat{\mathbf{x}}_0$, some starting random noise $\boldsymbol{\epsilon} \sim \mathcal{N}(0, I)$ is sampled, and the denoising procedure previously demonstrated is iteratively applied over a linear sigma-schedule.

# ARCHITECTURES

## 4.1 OUR `a-unet` LIBRARY

### 4.1.1 *Background of U-Net*

Diffusion models are commonly implemented with U-Nets [13], a type of convolutional architecture originally developed for image segmentation. U-Nets consist of an encoder network and a decoder network, connected by a series of skip connections that allow the model to learn and preserve fine details at multiple resolutions. The original architecture used 2D convolutions to exploit the spatial structure of images, but in our case we will adapt it for 1D convolutions in order to process raw multichannel signals. This allows for better use of the inductive biases of waveforms and greater speed.

The U-Net architecture has evolved over time, resulting in modern versions that incorporate numerous enhancements and improvements, including: new skip connections, convolutional blocks, attention blocks [7, 15], and several ways to change the behavior of the network by providing feature vectors or embeddings as context.

In order to deal with different variations, we provide `a-unet` a library that includes a toolbox of hackable building blocks to build a variety of U-Nets. The goal of `a-unet` is to provide the right level of abstraction for eventual extensions of the architecture, without requiring to alter the basic template so that experimentation and iteration speed can be increased.

### 4.1.2 *U-Net Block*

In order to build a generic U-Net block, it is necessary to identify the core components of the architecture (illustrated in Figure 7).

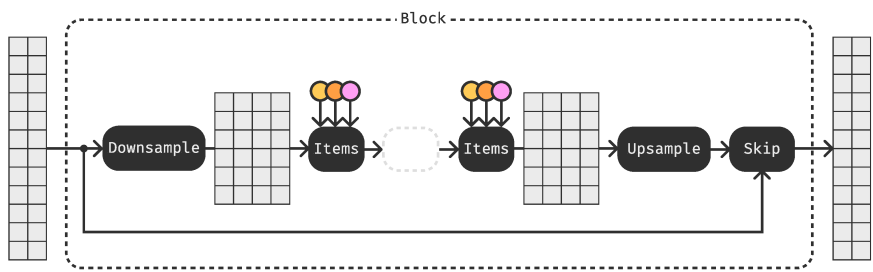

Figure 7: U-Net block

These include a *downsampling* block that simultaneously reduces the resolution and number of channels of the input (typically implemented with a single convolution), a stack of customizable processing *items* (see subsection 4.1.3 for details), an inner block that may contain another instance of the block recursively, a second stack of processing *items* that typically mirrors the first stack, an *upsampling* block that reverses the effects of the downsampling (typically implemented with a single transposed convolution), and a *skip* block that merges the skip connection using some operation.

Furthermore, we select 3 possible types of conditioning contexts that can be injected in the processing items, namely: a feature-vector based conditioning  typically used with diffusion to provide the noise level, an embedding based conditioning  that injects multiple embedding vectors as context, typically used for text/CLIP-embedding based conditioning, and lastly a channel-based conditioning  used to inject entire stacks of channels in the block. Depending on the task, we might a different combination of conditioning methods.

All described characteristics can be defined and customized using the following block:

```
from a_unet.apex import Block
block = Block(
    dim=1,
    in_channels=2,
    channels=4,
    factor=2,
    items=[...],
    # Optional
    items_up=[...],
    downsample_t=...,
    upsample_t=...,
    skipt_t=...,
    inner_block=...
)
```

This is a building block for a U-Net, where we can customize the number of input/ouput channels (`in_channels`), the number of `channels` post-downsampling, and the downsampling `factor`. The `items` list will contain the different items that will be duplicated after the inner block. Optionally, we can change the type of skip connection, downsampling and upsampling operations. The `inner_block` can be another instance of `Block` to recursively nest multiple blocks.

Since a U-Net is usually composed of multiple nested blocks where the number of `in_channles` of the inner block must match the number of `channels` of the outer block, we provide `XUnet` as a glue class, and `XBlock` as a template class for `Block` to make this process more convenient and automated:

```
from a_unet.apex import XUNet, XBlock
unet = XUNet(
```

```
        dim=1,
        in_channels=2,
        blocks=[
            XBlock(channels=4, factor=2, items=[...]),
            XBlock(channels=8, factor=2, items=[...]),
            XBlock(channels=16, factor=2, items=[...]),
        ],
        skip_t=...,
)
```

This is also very helpful to define generic properties like the type of skip connection `skip_t` in the `XUNet`, that will in turn automatically forwarded to all blocks. Parameters can also be provided to a specific `XBlock` to override the defaults.

### 4.1.3 Items

Items are the core processing units of the U-Net at each resolution. Out of the box we provide, a `ResnetItem` (R) as a convolutional processing unit, a `ModulationItem` (M) to apply modulation of the different channels provided a feature vector 🟠, an `AttentionItem` (A) for self-attention processing unit between region vectors, a `CrossAttentionItem` (C) for cross-attention between region vectors and a provided set of embedding vectors 🟠, a `FeedForwardItem` (F) for MLP like processing of region vectors, and a `InjectItem` (I) for injecting a set of provided external channels 🟣.

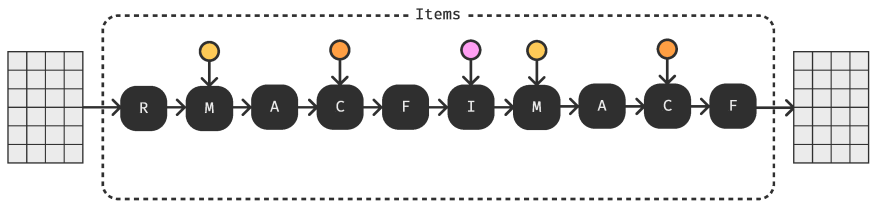

Figure 8: U-Net items

The example combination from Figure 8, or any other combination, can be easily added to `Block` or `XBlock` as follows:

```
from a_unet.apex import (
    XBlock
    ResnetItem as R,
    ModulationItem as M,
    AttentionItem as A,
    CrossAttentionItem as C,
    FeedForwardItem as F,
    InjectChannelsItem as I
)
stack = [M, A, C, F]
block = XBlock(..., items=[R] + stack + [I] + stack)
```

Additional customized items can be easily included without altering the template code, making experimentation very simple.

### 4.1.4 *Plugins*

Plugins are used to augment the U-Net model with additional functionality. It's often the case that we have to wrap the U-Net model with some pre- and post-transformation, or that we have to alter or augment the inputs provided to the U-Net. In order to maintain a modular structure, plugins can be used to directly modify the U-Net type without having to change the model code.

#### 4.1.4.1 *Time Conditioning Plugin*

The time conditioning plugin is used to convert a floating point value to a conditioning feature vector ◯, this is useful during diffusion to provide the current noise level, or timestep. To obtain the time feature vector from a floating point value, a learned weight is multiplied by the time information to get a frequency vector that is then processed using a pair of sin and cos to get Fourier features. The Fourier features are then transformed to a learned feature vector of the desired size by a stack of MLPs. This function can be easily added to the base U-Net as:

```
UNetWithTime = TimeConditioningPlugin(UNet)
```

This extends the U-Net with an additional `time` parameter, which can be one or more floating point values of each batch element.

#### 4.1.4.2 *Embedding Classifier Free Guidance Plugin*

Classifier free guidance is a method proposed in [4]. We provide a `ClassifierFreeGuidancePlugin` used to increase the conditioning strength of the provided embedding ◯. During training, the embedding is masked with a fixed (learned) embedding with a small probability in order to ensure that the network is able to generate realistic output without access to any conditioning information. During inference, the network is called twice, once with the conditioning embedding to get $\hat{y}_e$, and once with the fixed embedding used as mask to get $\hat{y}_m$. A scaling factor `embedding_scale` ($\lambda$) is then used to guide the network to produce an output that gives more or less importance to the conditioning embedding compared to the masked embedding:

$$\hat{y} = \hat{y}_m + (\hat{y}_m - \hat{y}_e) \cdot \lambda \tag{22}$$

This plugin can be easily used by augmenting the U-Net as:

```
UNetCFG = ClassifierFreeGuidancePlugin(
    net_t=UNet,
    embedding_max_length=64
)
```

Later the new `UNetCFG` model can be called with the additional parameter `embedding_mask_proba` to probabilistically mask a batch of embeddings during training (e.g. a value of 0.1 will mask 10% of the embeddings with a fixed embedding of length `embedding_max_length`), or with an `embedding_scale` parameter during inference, to call the U-Net twice with and without masking, and apply the scaling factor. In both cases, the `embedding` parameter must be provided as well.

### 4.1.4.3 *Text Conditioning*

The text conditioning plugin augments the U-Net embedding conditioning information  with a learned text embedding from a frozen pretrained language model. By default, the T5-base transformer model from [10] is used if no embedder is provided.

```
UNetWithText = TextConditioningPlugin(
    net_t=UNet,
    embedder=T5Embedder()
)
```

This adds an additional `text` field to the U-Net forward method that automatically extends the embedding with text embeddings from a pretrained language model.

## 4.2 OUR audio-encoders-pytorch LIBRARY

The autoencoder component has a similar structure to the U-Net, with a few changes: (1) to make it an autoencoder no skip connections will be used, (2) no attention blocks will be used to make it generic to any input sequence length (3) no conditioning blocks will be applied. We open-source the autoencoder library `audio-encoders-pytorch` (AEP) as a separate library from `a-unet`. AEP includes both encoders and decoders, and a set of bottlenecks that can be used to normalize the latent space, namely (1) a variational bottleneck in the style of VAEs [5], (2) a simple tanh bottleneck, (3) a quantizer bottleneck, similar to the one proposed by VQ-VAEs [9]. Furthermore, we propose two encoders that encode spectrograms channelwise into a 1D latent, namely a ME1d (magnitude spectrogram only encoder), or MelE1d (mel spectrogram encoder), both compatible with the different bottlenecks.

# MODELS 5

## 5.1 OVERVIEW

In this section we describe various diffusion models and their underlying structures. We investigate various diffusion models that serve different purposes and functions, including upsampling and autoencoding. Although these models may have distinct applications, they are ultimately utilized with the goal of audio generation. All of the different models are implemented using variations and combinations of the previously described achitectures (i.e. U-Nets and auto-encoders). The models proposed are implemented in the `audio-diffusion-pytorch` (ADP) library.

## 5.2 DIFFUSION UNCONDITIONAL GENERATOR

The diffusion generator is the simplest model we propose to synthetize unconditional audio and is implemented with a single 1D U-Net.

### 5.2.1 *Motivation*

The unconditional diffusion model is a good starting point to test the overall quality of the particular architecture and diffusion method used. It doesn't include any type of conditioning, making the dataset and training procedure very simple, and at the same time can give a good idea of the generation quality.

### 5.2.2 *Method*

The diffusion generator takes a raw high-quality stereo audio source as input from the datasets, that is then corrupted to a random noise level based on the chosen diffusion method. Using a U-Net, the generator then predicts the output, which may be the denoised input or a value that is used to compute the denoised input, depending on the type of diffusion method employed. The noise level (usually called *time* or $\sigma$) is provided as conditioning to the network thorugh as an encoded feature vector  to tell the network how much noise must be removed from the provided input. For the diffusion generator neither the embedding conditioning  nor the cross attention blocks are used.

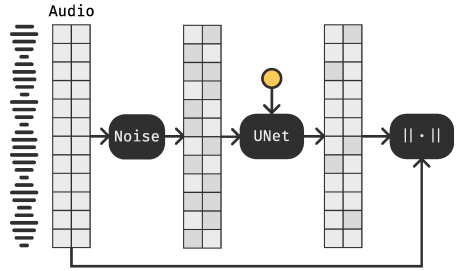

Figure 9: Diffusion model training

During inference, a random vector with the same shape as a training audio sample is sampled and the U-Net is iteratively invoked with varying noise levels to generate a new, plausible sample from the data distribution.

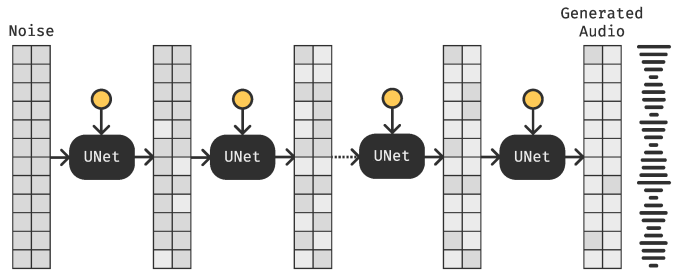

Figure 10: Diffusion model inference

### 5.2.3 *Diffusion Method*

We evaluated the performance of the proposed model with different diffusion methods. Out of the box, the model demonstrated good results with DDPM diffusion, but required approximately 200 sampling steps with a basic sampler during inference to generate reasonable results. If using k-diffusion and proper configuration of hyperparameters, the model struggles to generate high-quality audio that has a high dynamic-range, even with advanced sampling methods. If properly loudenss-normalized, we can get good results using around 10x less steps than DDPM diffusion, but we lose the capacity of producing samples that vary in loudness level. We found v-diffusion to be the most favorable method, robust to unnormalized audio and still reasonably fast to sample from, with around 50 sampling steps produces good results using DDIM sampling. In addition to that, v-diffusion requires tuning fewer hyperparameters, giving a good balance between simplicity, speed, and sample quality.

5.2.4 *Transforms*

Independently of the diffusion method used, this model without any addition struggles to generate more than a few second of sound. If the raw waveform is provided to the network the initial convolutional blocks of the U-Net will have to process huge samples, e.g. even a single second of high-quality 48kHz audio requires 48000 values to be processed by the first convolutional block. This can be a speed issue if the audio is not downsampled quickly enough in the U-Net, as the inefficiency will compound over the number of sampling steps of the diffusion process. In addition to that, if attention blocks are used, we will have to downsample enough to make sure that the number of timesteps to be in the range of 1024 or 2048 values. Exceeding that will slow down self-attention drastically due to the $n^2$ computational complexity for sequence length $n$. Hence, a lot of downsampling is required with long audio samples if we want to satisfy these criteria.

To mitigate the challenges mentioned earlier, we investigate the use of various methods and audio transforms to convert the raw audio source into a representation that reduces the temporal dimension in exchange for additional channels.

5.2.4.1 *Patching*

The first transform is patching, proposed originally for the image domain in [6]. We adapt patching to the 1D domain, where the idea is to group sequential time steps into chunks, that will then be transposed to channels. Given a patch size $p$, the length $t$ is reduced by $\frac{t}{p}$ where the number of channels increases to $c \cdot p$, at the end of the U-Net processing the channels are unchunked back to the full length. We found patching to give drastic speedups, almost at a factor of $p$ for $p = 2, 4, 8, 16, 32, ...$, allowing to train models with much longer audio sources. However, even if the audio generation quality almost matches the non-patched version, audible aliasing is present with all factors. This drawback is likely due to the repeated unchunking process, which will have a repeating structure, creating a high-frequency sine wave in the signal. Furthermore, we found that patching with $p \geqslant 64$ started to degrade quality, probably due to some capacity constraint in the channel dimension. We can think of patching as a deterministic auto-encoding process, with a downsampling factor of $p$.

5.2.4.2 *STFT*

The second transform is the previously introduced STFT. We use the common setting of 1024 num fft and window length with 256 hop size. By wrapping the U-Net with STFT and iSTFT the transform downsamples the length of the audio by 1024 while equally increas-ing the channel count. STFT is implemented with the Fast-Fourier Transform, hence it's efficient to apply. No normalization is required on the spectrogram, since the diffusion loss will still be applied on the reconstructed wave. This method gives great speedups thanks to the large downsampling, but similarly to patching suffers from degradation in quality compared to the raw wave representation. Perceptible noise is present in the generations both when transforming to magnitude+phase, or when using real+complex.

5.2.4.3 *Learned Transform*

Lastly, we propose a learned transformation with a single convolutional and transposed-convolutional block at the start and respectively end of the U-Net. The transform consists in using a large kernel size and stride of 64. This will down-sample the original signal in a single step, trading off small amounts of speed from the deterministic patching or FFT implemented STFT. However, since it's a convolutional method, we can choose the number of channels and increase it to a larger value (e.g. 128, double the kernel size and stride) than used during patching, giving more capacity to be resilient to artifacts. At the same time, we can use ideas from STFT and have large overlapping windows with learned kernels instead of fixed sine/cosine waves (e.g. kernel size 128, stride 64, 64 channels, with padding to preserve dimension), which can help to overcome aliasing. We found this to be the best quality/speed tradeoff method of pre-transforming audio.

5.2.5 *Usage*

The diffusion generation model proposed is constructed by first adding the `LTPlugin` to the default U-Net `UNetV0`. This plugin wraps the U-Net with the previously described learned transform. After that, we have to provide the U-Net type to the `DiffusionModel` class which is responsible for constructing the U-Net, the diffusion training method (by default V-Diffusion), and the diffusion sampler (by default DDIM).

```
from audio_diffusion_pytorch import DiffusionModel, UNetV0,
    LTPlugin, VDiffusion, VSampler

UNet = LTPlugin(
    UNetV0, num_filters=128, window_length=64, stride=64
)

model = DiffusionModel(
    net_t=UNet,
    in_channels=channels,
    channels=[256, 256, 512, 512, 1024, 1024],
    factors=[1, 2, 2, 2, 2, 2],
    items=[2, 2, 2, 2, 4, 4],
```

```
        attentions=[0, 0, 0, 0, 1, 1],
        attention_features=64,
        attention_heads=12,
        diffusion_t=VDiffusion,
        sampler_t=VSampler
)
```

This model can be easily used to get the diffusion loss for training (which automatically applies the entire diffusion process) or to sample a new element provided the starting noise.

```
# Training
x = torch.randn(1, 2, 2**21) # [batch, channels, length]
loss = model(x)
# Sampling
noise = torch.randn(1, 2, 2**21)
sample = model.sample(noise=x, num_steps=50)
```

### 5.2.6 *Evaluation*

We found that it's important for quality to have a single non-downsampled block at the start to process the transformed audio at full resolution. Furthermore, attention blocks are crucial for temporal consistency of the generated audio, but can only be applied after the original waveform is down sampled to around 1024-2048 length. For example, if the original audio has length $2^{19}$ (i.e. ~11s at 48kHz), we downsample by $64 = 2^6$ in the learned transform, and by $2^3$ in the 4 blocks before the first attention block, hence the context length of the first attention blocks will be in the desired range of $2^{10} = 1024$.

This model can generate high quality audio over tens of seconds, possibly more depending on the speed requirements. In general, a larger set of initial convolutional/resnet blocks (closer to the waveform) will result in better audio quality, at the cost of generation speed.

We found that the architecture is able to generalize to longer samples than it was trained on, if attention blocks are used. The samples maintain good long-context awareness even when doubling or more the training length. Note that this increases the attention context size and hence needs to be considered for before training.

### 5.3 TEXT-CONDITIONAL DIFFUSION

### 5.3.1 *Motivation*

We used text as a mean of conditioning for several reasons. In Imagen [15] it has been shown that pretrained and frozen language models can be successfully applied to condition the diffusion process to generate images matching a textual description, and that by increasing the size of the language model will result in an improved text-image matching. This hints to the fact that a similar method might also work well with audio. Furthermore, text is a useful conditioning method as multiple things can be encoded in text (text-to-music, text-to-speech), making the interface more generic and easy to use.

5.3.2  *Method*

The previous unconditional model can be easily adapted to allow for text conditioning by adding cross attention blocks at different resolutions of the U-Net. Following [15], we use a frozen and pretrained T5 transformer encoder to encode the textual representation into an embedding, which is used to condition the diffusion model. For music generation, we train on metadata, including the title of the song, author, album, genre, year of release. Since each training chunks lasts 44s each, we append a string indicating the which chunk is being trained on and how many total chunks the song is made of (e.g. *1 of 4*), this allows during inference to select the region of interest, most commonly start (*1 of N*) or end (*N of N*). To make the conditioning more robust, we shuffle the list of metadata and drop each element with a probability of 0.1. Furthermore, 50% of the times we concatenate the list with spaces and the other 50% of the times we use commas to make the interface more robust.

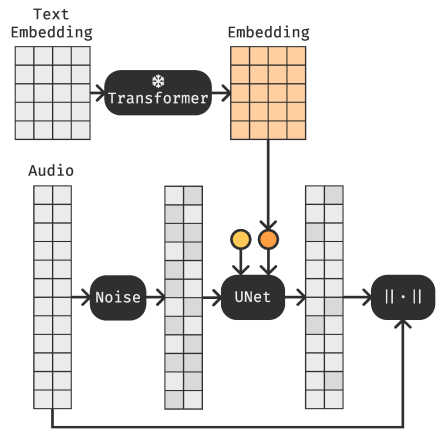

Figure 11: Text-conditional diffusion model training

To increase the strength of the text conditioning, we apply classifier-free guidance [4]. During training, the text embedding is dropped 10% of the times, in favor of a fixed learned embedding. Text conditioning with T5 and CFG can be easily added to the model with the following additional attributes:

```
model = DiffusionModel(
    ...
    cross_attentions=[1, 1, 1, 1, 1, 1],
```

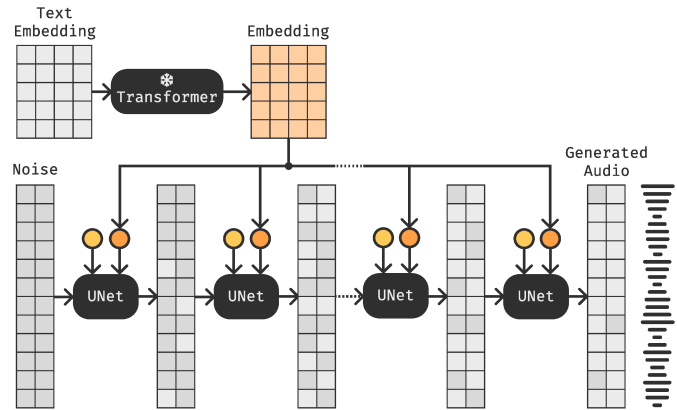

Figure 12: Text-conditional diffusion model inference

```
    use_text_conditioning=True,
    use_embedding_cfg=True,
    embedding_features=768,
    embedding_max_length=64,
)
# Training
loss = model(x, text=['my text'], embedding_mask_proba=0.1)
# Sampling
sample = model.sample(
    noise,
    text=['my text'],
    num_steps=50,
    embedding_scale=5.0,
)
```

Which internally extend the U-Net with cross attention blocks, the text conditioning plugin, and the classifier free guidance plugin.

### 5.3.3 *Evaluation*

We found text conditioning to work well to match audio with the textual description, especially using the genre of the song or more generic words that are found in titles. We also tried text-to-speech (TTS), but found that the model is able to mumble a few words with good audio quality, but is not able to have a correct ordering. This is a common problem in TTS that is usually solved with the help of an alignment algorithm that provides additional information about the text-audio positioning, or by training transformers autoregressively.

## 5.4 DIFFUSION AUTO-ENCODERS WITH LATENT DIFFUSION

### 5.4.1 *Motivation*

Patching, STFT, and learned transforms can be used to reduce the input length during the diffusion process. Those approaches are advantageous if we want to train a single model end-to-end, however, this is suboptimal since the waveform is expanded to its original full-length shape multiple times during sampling, slowing down the process.

A more appropriate way would be to first encode the waveform, then do the diffusion loop in the compressed representation, never expanding it to the full waveform until the end of the loop. This is the idea proposed in [12] (latent diffusion), where a variational autoencoder is first used to compress images by a few factors to a smaller latent space, and later diffusion is applied to that latent. By compressing the audio before applying diffusion, we can drastically speed up the diffusion sampling procedure, making an important case for an efficient and good quality autoencoder.

### 5.4.2 *Method*

There are different ways to implement the autoencoder, however an important property is that we must be able to apply the diffusion process to its latent space, hence some sort of normalization is required to make sure the values are in the range $[-1, 1]$. Furthermore, the autoencoder should compress as much as possible without a significant loss in quality. The smaller the latent, the faster will be the inner diffusion model to process and generate.

We experimented with different autoencoders, and found that directly compressing the waveform can only provide around 2x-4x compression without a significant loss in quality. On the other hand, as we have discussed in the representation section, compressing magnitude or mel spectrograms can provide much higher compression rates. The downside is that the spectrogram requires a model (vocoder) to reconstruct the original waveform, even from a non-compressed state.

In this work, we propose to use a *magnitude* diffusion autoencoder, an encoder (ME1d) first encodes the waveform into a magnitude spectrogram which is then encoded into a latent compressed 64x compared to the original waveform, and later uses a diffusion model to reconstruct the waveform conditioned on the latent, acting both as a deterministic compressing encoder and a diffusion vocoder at the same time. In order to make sure the latent space is normalized, we use a tanh function on the bottleneck. Since the decoding/vocoding process is a diffusion model, the waveform can be quickly reconstructed from the latent by using a small step count, if instead a more accurate reconstruction is desired a higher step count is required. To make sure the diffusion autoencoder is generic to any waveform length, we remove attention blocks and use a convolutional-only architecture. The latent is injected directly in the U-Net at the right resolution by concatenating the channels.

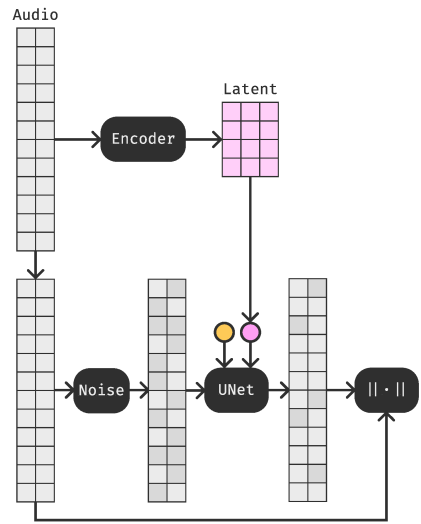

Figure 13: Diffusion autoencoder training

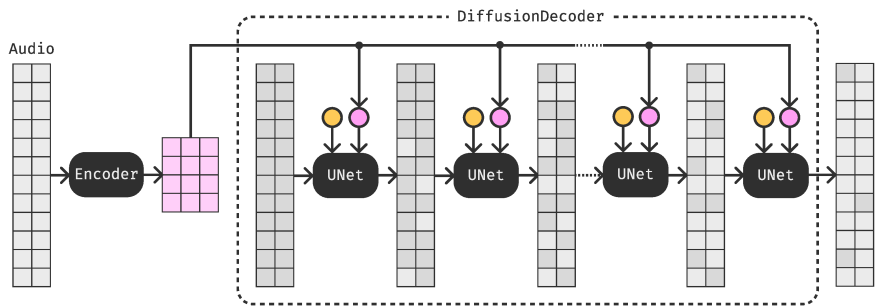

Figure 14: Diffusion autoencoder inference

To get the final model, we apply a cascading diffusion generator to generate the latent with text conditioning, in the style of latent diffusion. Since the representation is dramatically compressed, we can train with much longer waveforms, and use the trained diffusion decoder to expand the representation back to waveform.

### 5.4.3 *Evaluation*

This architecture provides a mediocre generation quality, faster than real-time generation speed on a single GPU, and large context length. It can generate music with consistent rythm over multiple minutes, and has a good text-audio binding.

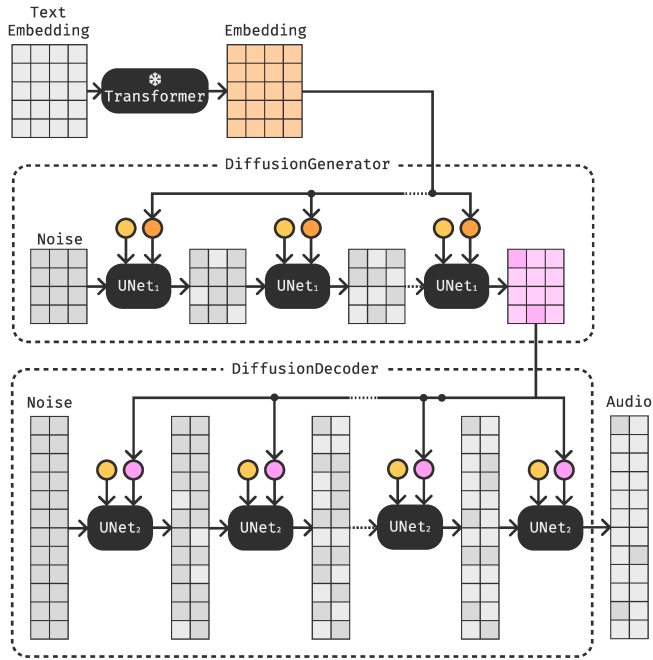

Figure 15: Two-stage diffusion generator with diffusion decoder

## 5.5 DIFFUSION UPSAMPLER

### 5.5.1 *Motivation*

The diffusion upsampler is used to increase the sampling rate of a provided waveform (e.g. from 3kHz to 48kHz). From the perspective of spectrograms, downsampling a waveform corresponds to setting to zero the top half of the grid (or image) starting at some frequency threshold, where upsampling corresponds to reconstructing the missing zeros. The model we propose, however, works directly on waveforms. Diffusion upsamplers can be seen as a specific case of diffusion autoencoders, where the encoding function is fixed to be the downsampling operation. Upsamplers can be used for different purposes: (1) as autoencoders on which to apply latent diffusion, i.e. generate a low sample rate audio with a primary model and later upsample a secondary upsampler model, (2) to increase the sample rate of existing low sample rate audio by training (learning from) high sample rate audio.

### 5.5.2 *Method*

The conditioning is applied by artificially upsampling the channels using interpolation, and appending them as additional context to the input of the diffusion model.

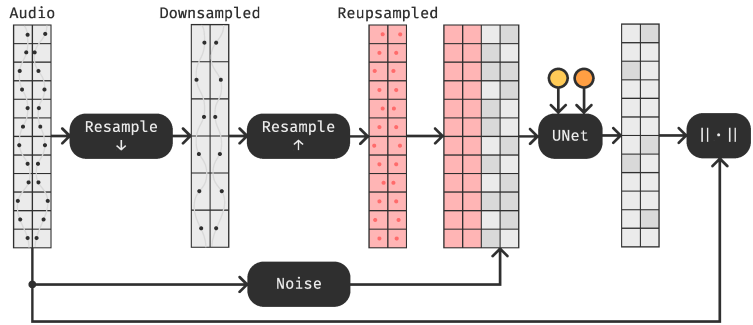

Figure 16: Diffusion upsampler training

We artificially interpolate the channels to make sure the length matches the output high sample channels and hence can be properly stacked.

During inference, we interpolate the low sample channels to match the high sample rate length and use the sampling process to reconstruct the missing details.

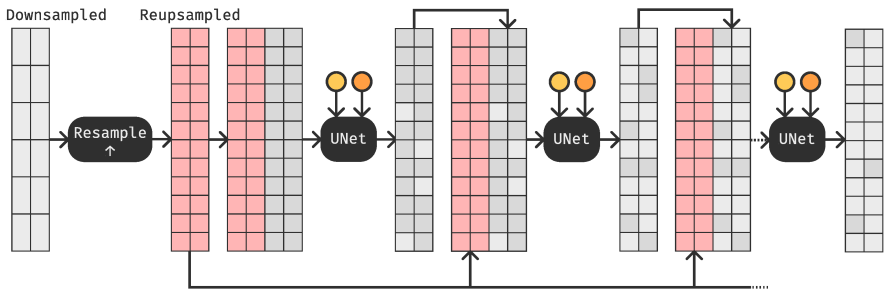

Figure 17: Diffusion upsampler inference

Optionally, additional context such as a text description in the form of an embedding ◯ can be provided as additional guidance to help the reconstruction, especially if upsampling from very low sample rates.

### 5.5.3 *Evaluation*

Depending on the complexity of the dataset, diffusion upsamplers can get very good results by upsampling anywhere between 16x and 64x. We found upsamplers to excel on speech data, as it's likely an easier modality. Similarly to other models, increasing the size (channel count or layers) of the initial convolutional blocks in the U-Net corresponds with higher quality, especially to generate high frequency wavefroms. Attention blocks and larger context lengths can help upsamplers to infer missing parts of the song by referring to audio that is further apart, similarly to generation models.

## 5.6 DIFFUSION VOCODER

### 5.6.1 *Motivation*

Mel-spectrogram perceptually balanced to the frequencies we perceive, making them an ideal representation for audio generation. However, properly turning a spectrogram back to a playable audio waveform is not a trivial task. Some iterative methods such as Griffin-Lim tend to produce artifacts, making the case for commonly used deep-learning based vocoders. Trained vocoders can produce very good results for mel-to-speech, but open-source implementations for high-quality 48kHz music vocoding are still lacking [1]. In the following section, we propose a simple adaptation that allows to turn our U-Net architecture with almost no change into high-quality music vocoder.

### 5.6.2 *Method*

The diffusion vocoder is trained by first converting the waveform to spectrogram, then flattening the spectrogram using a transposed 1d convolution back to its waveform shape. Similarly to the upsampler, we stack the additional channels on the input channels of the U-Net.

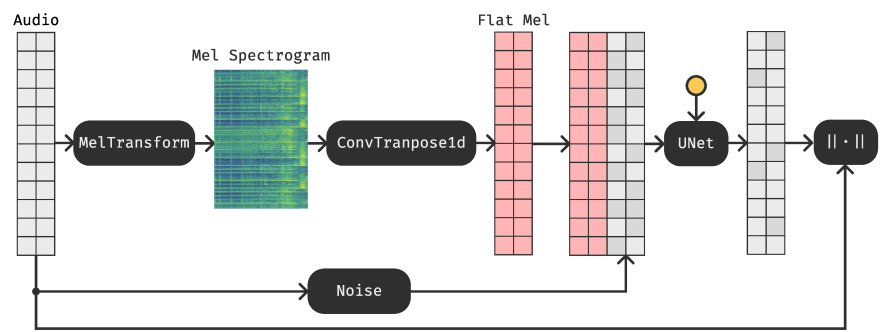

Figure 18: Diffusion vocoder training

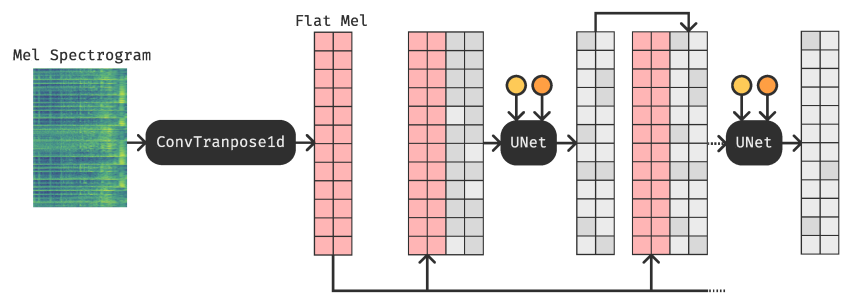

Figure 19: Diffusion vocoder inference

---

[1] https://github.com/NVIDIA/BigVGAN

In order to flatten the spectrogram, we have to match the configuration of the STFT used to obtain the spectrogram, with the configuration of the 1d transposed convolution. The key insight is that the STFT operation can be viewed as a 1D convolution with large kernel sizes (or window size) of sine and cosine waves, which is then merged in-place using the absolute value, and later mel-scaled. The mel-scaling doesn't alter the temporal positioning, only the frequency (or channels) of the spectrogram. Hence, if we set large kernel sizes equivalent to the STFT window length, strides equivalent to the STFT hop-length, and proper padding, the transposed convolution will focus on the same context region of the waveform used to obtain the spectrogram. Similarly, we will set the input channels of the transposed convolution to match the number of channels used for the mel-spectrogram, and the output channels to 1. Stereo audio is decoded by batching. We used a window-length/kernel-size of 1024 and hop-length/stride of 256, similarly to popular vocoders we used 80 mel-spectrogram channels. With this configuration, the spectrogram has a default 3.2x compression factor over the initial waveform.

### 5.6.3 *Evaluation*

This model can produce high quality waveform, as with other models, a good reconstruction of high-frequencies requires more convolutional blocks towards the start of the U-Net. Moreover, we hypothesize that increasing the number of mel-channels might increase quality for two reasons: first, mel-spectrogram would compress less information out of the initial waveform, and second, the transposed convolution would have more channels to flatten the spectrogram and hence more capacity.

## 5.7 TRAINING INFO

### 5.7.1 *Data*

We trained all of our models on a 2500h mix of audio at 48kHz. In the text-based model, we used metadata such as title, genre, album and artist as conditioning information. For the autoencoder, upsampler, vocoder, we trained on random crops of length $2^{18}$ (~5.5s at 48kHz). For the long-context text-conditional audio generation model, we trained on fixed crops of length $2^{21}$ (~44s at 48kHz), using the crop index as additional conditioning information.

### 5.7.2 *Training*

We trained all of our models with AdamW, using a learning rate of $10^{-4}$, $\beta_1 = 0.95$, $\beta_2 = 0.999$, $\epsilon = 10^{-6}$, and wight decay of $10^{-3}$. For all models, we used an exponential moving average with $\beta = 0.995$ and power of $0.7$. We trained all models for around 1M steps with a batch size of 32, this takes approximately 1 week on a single A100 GPU for the largest, text-conditional model.

# 6

## FUTURE WORK

While our models can have a good generation quality on short few-second segments, or a good structure with longer segments, training an efficient model with both high quality and long context remains an open problem. A few promising future modelling approaches that need more experimentation include: (1) train diffusion models using perceptual losses on the waveforms instead of L2, this might help to decrease the initial size of the U-Net, as we wouldn't have to process non-percieveable sounds, (2) stack multiple upsamplers to generate a song top-down from low-sample rates to high sample rates, (3) improve the quality of the diffusion autoencoder by using mel-spectrograms instead of magnitude spectrograms as input, (4) other types of conditioning which are not text-based might be useful to navigate the audio latent space, which is often hard to describe in words - DreamBooth-like models [14] could be used to assign symbols to sounds, (5) compress mel-spectrograms to a quantized representation with diffusion autoencoders to allow for high compression ratios and later train an autoregressive transformer on top of that.

Other simpler improvements on the current models include: (1) increase the training data from 2k hours to 60k-100k hours, (2) use more sophisticated diffusion samplers to get higher quality for the same number of sampling steps, (3) for text-based models, use larger pretrained language to obtain embeddings, which has been shown to be very important for quality in [15].

# 7
## CONCLUSION

Generating high-quality audio efficiently is a challenging task as it involves the generation of numerous values to accurately represent the sound waves, especially when aiming for high-fidelity stereo sound at a sample rate of 48kHz. In this work, we proposed different methods and models to generate high quality audio from a textual description. From models targeting long-context audio with an emphasis on structure, short-context with an emphasis on quality, to other useful models such as the diffusion upsampler and vocoder. We introduced a new method that utilizes text-conditional diffusion models based on 1D U-Nets, allowing for the generation of multiple minutes of 48kHz audio on a single consumer GPU. Furthermore, we have provided a collection of open-source libraries to streamline future research, including potential improvements in audio autoencoders and diffusion models.

# BIBLIOGRAPHY

[1] Zalán Borsos, Raphaël Marinier, Damien Vincent, Eugene Kharitonov, Olivier Pietquin, Matt Sharifi, Olivier Teboul, David Grangier, Marco Tagliasacchi, and Neil Zeghidour. *AudioLM: a Language Modeling Approach to Audio Generation*. 2022. eprint: arXiv:2209.03143.

[2] Prafulla Dhariwal, Heewoo Jun, Christine Payne, Jong Wook Kim, Alec Radford, and Ilya Sutskever. *Jukebox: A Generative Model for Music*. 2020. eprint: arXiv:2005.00341.

[3] Jonathan Ho, Ajay Jain, and Pieter Abbeel. "Denoising diffusion probabilistic models." In: *Advances in Neural Information Processing Systems* 33 (Dec. 2020), pp. 6840–6851.

[4] Jonathan Ho and Tim Salimans. *Classifier-Free Diffusion Guidance*. 2022. eprint: arXiv:2207.12598.

[5] Diederik P Kingma and Max Welling. *Auto-Encoding Variational Bayes*. 2013. eprint: arXiv:1312.6114.

[6] Troy Luhman and Eric Luhman. *Improving Diffusion Model Efficiency Through Patching*. 2022. eprint: arXiv:2207.04316.

[7] Ozan Oktay et al. *Attention U-Net: Learning Where to Look for the Pancreas*. 2018. eprint: arXiv:1804.03999.

[8] Aaron van den Oord, Sander Dieleman, Heiga Zen, Karen Simonyan, Oriol Vinyals, Alex Graves, Nal Kalchbrenner, Andrew Senior, and Koray Kavukcuoglu. *WaveNet: A Generative Model for Raw Audio*. 2016. eprint: arXiv:1609.03499.

[9] Aaron van den Oord, Oriol Vinyals, and Koray Kavukcuoglu. *Neural Discrete Representation Learning*. 2017. eprint: arXiv:1711.00937.

[10] Colin Raffel, Noam Shazeer, Adam Roberts, Katherine Lee, Sharan Narang, Michael Matena, Yanqi Zhou, Wei Li, and Peter J. Liu. *Exploring the Limits of Transfer Learning with a Unified Text-to-Text Transformer*. 2019. eprint: arXiv:1910.10683.

[11] Aditya Ramesh, Prafulla Dhariwal, Alex Nichol, Casey Chu, and Mark Chen. *Hierarchical Text-Conditional Image Generation with CLIP Latents*. 2022. eprint: arXiv:2204.06125.

[12] Robin Rombach, Andreas Blattmann, Dominik Lorenz, Patrick Esser, and Björn Ommer. *High-Resolution Image Synthesis with Latent Diffusion Models*. 2021. eprint: arXiv:2112.10752.

[13] Olaf Ronneberger, Philipp Fischer, and Thomas Brox. *U-Net: Convolutional Networks for Biomedical Image Segmentation*. 2015. eprint: arXiv:1505.04597.

[14] Nataniel Ruiz, Yuanzhen Li, Varun Jampani, Yael Pritch, Michael Rubinstein, and Kfir Aberman. *DreamBooth: Fine Tuning Text-to-Image Diffusion Models for Subject-Driven Generation*. 2022. eprint: `arXiv:2208.12242`.

[15] Chitwan Saharia et al. *Photorealistic Text-to-Image Diffusion Models with Deep Language Understanding*. 2022. eprint: `arXiv:2205.11487`.

[16] Tim Salimans and Jonathan Ho. *Progressive Distillation for Fast Sampling of Diffusion Models*. 2022. eprint: `arXiv:2202.00512`.

[17] Jascha Sohl-Dickstein, Eric A. Weiss, Niru Maheswaranathan, and Surya Ganguli. *Deep Unsupervised Learning using Nonequilibrium Thermodynamics*. 2015. eprint: `arXiv:1503.03585`.

[18] Jiaming Song, Chenlin Meng, and Stefano Ermon. *Denoising Diffusion Implicit Models*. 2020. eprint: `arXiv:2010.02502`.

[19] Ashish Vaswani, Noam Shazeer, Niki Parmar, Jakob Uszkoreit, Llion Jones, Aidan N. Gomez, Lukasz Kaiser, and Illia Polosukhin. *Attention Is All You Need*. 2017. eprint: `arXiv:1706.03762`.